\documentclass[fleqn,10pt]{wlscirep}
\usepackage[utf8]{inputenc}
\usepackage[T1]{fontenc}
\usepackage{amsmath}
\usepackage{graphicx}
\usepackage{url}
\usepackage{cite}
\usepackage{booktabs}
\usepackage{longtable}

\title{Extratropical Atmospheric Circulation Response to ENSO in Deep Learning Pacific Pacemaker Experiments}

\author[1]{Zhanxiang Hua}
\author[1,*]{Christina Karamperidou}
\author[2]{Zilu Meng}
\affil[1]{Department of Atmospheric Sciences, University of Hawai\textquoteleft i at M\={a}noa, Honolulu, HI, USA}
\affil[2]{Department of Atmospheric and Climate Science, University of Washington, Seattle, WA, USA}

\affil[*]{ckaramp@hawaii.edu}

\begin{abstract}
Coupled atmosphere-ocean deep learning (DL) climate emulators are a new frontier but are known to exhibit weak ENSO variability, raising questions about their ability to simulate teleconnections. Here, we present the first Pacific pacemaker (PACE) experiments using a coupled DL emulator (DLESyM) to bypass this weak variability and isolate the atmospheric response to observed ENSO forcing. We find that while the emulator realistically captures internal atmospheric variability, it produces a significantly amplified forced teleconnection response to ENSO. This amplified response leads to biases in simulating extremes, notably an overestimation of atmospheric blocking frequency and duration with the underestimation of peak intensity. Our findings underscore that coupled DL climate models require in-depth and physically-grounded validation, analogous to traditional numerical models, to build confidence in their use for physical climate analysis.

\end{abstract}
\begin{document}

\flushbottom
\maketitle
\thispagestyle{empty}

\section*{Introduction}

The El Niño–Southern Oscillation (ENSO) is the dominant mode of interannual climate variability, exerting far-reaching impacts on global atmospheric circulation, precipitation, and temperature patterns \cite{Trenberth1997, McPhaden2006}. Through atmospheric teleconnections, tropical Pacific sea surface temperature (SST) anomalies and the resulting shifts in tropical convection excite poleward-propagating Rossby waves, giving rise to characteristic extratropical geopotential height responses such as the Pacific–North American (PNA) pattern during boreal winter \cite{Wallace1981}. These teleconnections underpin seasonal predictability and influence extreme events including droughts, floods, and midlatitude atmospheric blocking \cite{Seager2015, Renwick1996,McKenna2023TheIO}.

Despite extensive progress, state-of-the-art coupled models continue to exhibit systematic ENSO biases, including weak amplitude, frequency errors, and underestimated extratropical teleconnection strength \cite{Bayr2018, Fasullo2020}. Such deficiencies complicate the attribution of teleconnection errors, since biases in tropical forcing, atmospheric process representation, or the background climatological state can each alter the structure and phase of extratropical responses \cite{Deser2017, Greatbatch2012}. For example, mean state biases such as misplaced jet streams or SST gradients are known to anchor waveguides, inducing shifts in the phase and strength of the ENSO–Z500 teleconnection pattern \cite{Watanabe2014}.

Pacemaker experiments were developed to address this challenge by constraining tropical SST variability while preserving coupled dynamics elsewhere \cite{Kosaka2013, Deser2017}. By nudging SST anomalies in a specific region (e.g., the equatorial Pacific) to observations, pacemaker experiments isolate the causal impact of observed SST variability on global circulation. This design reduces contamination from tropical SST biases and enables systematic evaluation of ENSO teleconnections, forced responses, and internal atmospheric variability \cite{Deser2010, Kosaka2013}. Pacemaker experiments have thus been employed to estimated uncertainties in North Pacific teleconnections and highlight state dependence of ENSO impacts on the Arctic and Europe \cite{Deser2017, Simpson2018}.

Meanwhile, recent advances in machine learning have yielded deep learning climate emulators that capture many of the spatial and temporal patterns reproduced by full dynamical models at a fraction of the computational cost \cite{Beucler2021, WattMeyer2023ACEAF,WattMeyer2024ACE2AL}. Atmosphere-only emulators, such as the Ai2 Climate Emulator (ACE2), can skillfully reproduce the spatial patterns of ENSO's atmospheric response when forced with observed historical sea surface temperatures \cite{WattMeyer2024ACE2AL}. However, since ENSO is an intrinsically coupled phenomenon, fully coupled emulators are necessary to study its dynamics and the atmospheric response to its forcing. The Deep Learning Earth System Model (DLESyM) represents a step in this direction, coupling a deep learning atmosphere with a simplified deep learning ocean model \cite{CresswellClay2024ADL}. DLESyM is trained on ERA5 reanalysis from 1979 to 2017 for both atmospheric variables and SST to minimize short-term forecast errors, yet it generates realistic long-term climate variability in free-running simulations. Despite capturing the main modes of variability, DLESyM suffers from a weak ENSO amplitude. This weakness presents a fundamental challenge for model validation. Even if the free-running model produces realistic extratropical circulation, it is unclear whether this realism stems from a correctly captured teleconnection to the tropics or from learned statistical associations with local variables. Therefore, to determine if the model has learned the teleconnection between tropical forcing and the extratropical response, it is necessary to isolate this pathway using the pacemaker framework.

To date, pacemaker experiments have not been performed with coupled deep learning climate emulators, despite their importance for addressing the fundamental scientific objective of simulating the tropical-extratropical interactions realistically. Only after validating this, can we leverage the emulators' unique advantages: huge ensembles enabled by computational efficiency \cite{meng2025deep}, transferability across climate applications, and the potential for gradient-based attribution for uncovering climate interactions. Applying pacemaker methodology to DLESyM offers a pathway to test whether constraining tropical Pacific SST anomalies can alleviate weak ENSO teleconnections, and to assess how background climatology influences extratropical responses.

Here we present the first, to our knowledge, application of the Pacific pacemaker (PACE) framework to a deep learning climate emulator. First, we conduct a pacemaker experiment where the tropical Pacific SSTs in DLESyM are nudged to observed values, analogous to the Community Earth System Model 2 (CESM2\cite{danabasoglu2020community}) Pacific Pacemaker (PACE) experiment \cite{Deser2017, Kosaka2013}. This setup allows us to bypass DLESyM's native weak ENSO forcing and directly assess the response of its atmospheric component to realistic tropical variability, addressing: (1) Does DLESyM's atmospheric component capture the primary spatial pattern and a realistic amplitude of the ENSO teleconnection when forced with observed tropical pacific SST anomalies? Furthermore, to disentangle the influence of the atmospheric model physics from that of the background climate state, we perform a second pacemaker experiment where DLESyM's atmosphere is forced using the mean-state climatology from CESM2 instead of its own. This allows us to investigate: (2) How do differences in background tropical Pacific climatology induce phase shifts in Z500 teleconnection patterns? Finally, by analyzing the outcomes of these forced simulations, we evaluate: (3) Can the pacemaker-forced DLESyM realistically simulate the modulation of extratropical extremes, specifically the frequency of atmospheric blocking, by ENSO?

By comparing these targeted simulations with the freerunning DLESyM and the benchmark CESM2 pacemaker ensemble, we provide the first systematic evaluation of ENSO teleconnections in a deep learning climate emulator under pacemaker forcing, thereby advancing the validation of deep learning models as physically consistent tools for climate science.

\section*{Results}

\begin{figure}[ht]
\centering
\includegraphics[width=\linewidth]{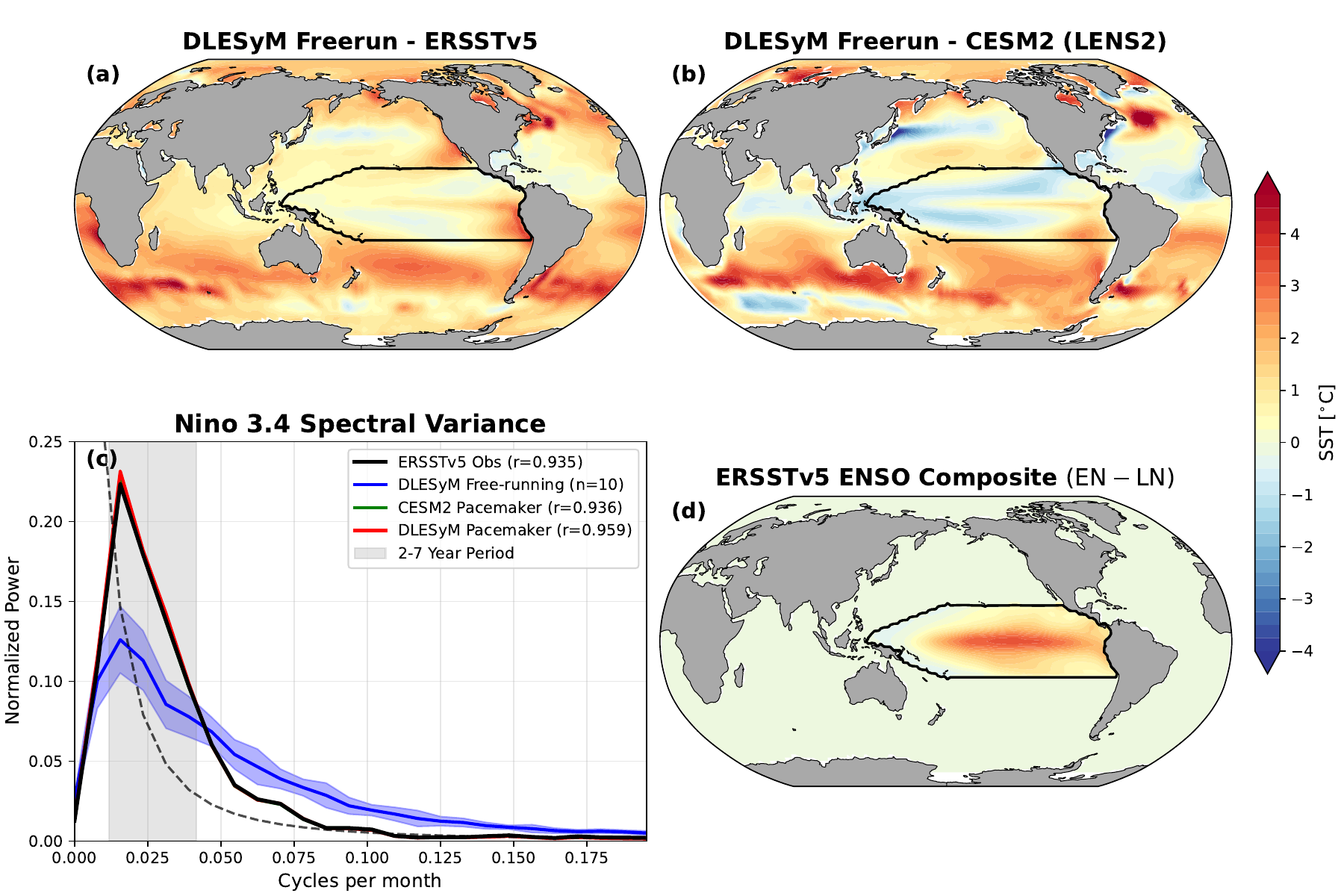}
\caption{(a–b) Climatological sea surface temperature (SST) differences of the DLESyM free-running simulation relative to (a) the ERSSTv5 observational dataset and (b) the CESM2 Large Ensemble (LENS2). The black contour indicates the pacemaker masking region. (c) Normalized spectral variance of the Niño~3.4 index following the methodology of Cresswell-Clay et al. 2025 \cite{CresswellClay2024ADL}. 
The correlation coefficient ($r$) denotes the lag-1 autocorrelation of each Niño~3.4 time series derived from the models and ERSSTv5. 
The dashed line represents the theoretical red-noise spectrum corresponding to the observed lag-1 autocorrelation ($r$). 
The gray shaded region indicates the 2–7~year ENSO band. 
For DLESyM, $n$ denotes the number of ensemble members, and the shaded envelope represents one standard deviation across the ensemble. 
(d) DJF ENSO (El~Niño $-$ La~Niña) composite of SST anomalies from ERSSTv5.
}
\label{fig:intro}
\end{figure}

DLESyM's climatological SST exhibits a warm bias in the equatorial Pacific cold tongue and Southern Ocean, and a cold departure in the Kuroshio Current extension region (Figure.~\ref{fig:intro}a,.~\ref{fig:intro}b) when compared to both Extended Reconstructed Sea Surface Temperature, Version 5 (ERSSTv5 \cite{huang2017extended}) and CESM2. This may be attributed to the fact that DLESyM is trained on ERA5 data from 1983 to 2016, a comparatively warmer period.
These baseline biases are important as they can influence the atmospheric response to remote forcing.

At interannual timescales, the free-running DLESyM confirms a limitation previously documented in deep learning emulators lacking explicit subsurface ocean dynamics: a substantially weakened ENSO amplitude. This is evident in the reduced spectral power within the canonical 2–7 year ENSO frequency band compared to observations (Fig.~\ref{fig:intro}c). This weak intrinsic ENSO variability necessitates the use of a pacemaker framework to robustly evaluate the model's atmospheric teleconnections. By nudging tropical Pacific SSTs to follow the observed historical record, we bypass the model's weak internal forcing and impose a realistic ENSO signal within the observed power spectrum (Fig.~\ref{fig:intro}c). The classic wintertime (DJF) ENSO SST anomaly pattern, shown as the \textbf{ENSO Composite}, is defined as the difference between all El~Niño and La~Niña events. This pattern serves as the target forcing, providing a tropical driver for the extratropical atmospheric response we investigate (Fig.~\ref{fig:intro}d).

\subsection*{Evaluating midlatitude response to ENSO in DLESyM}
\begin{figure}[!ht]
\centering
\includegraphics[width=\linewidth]{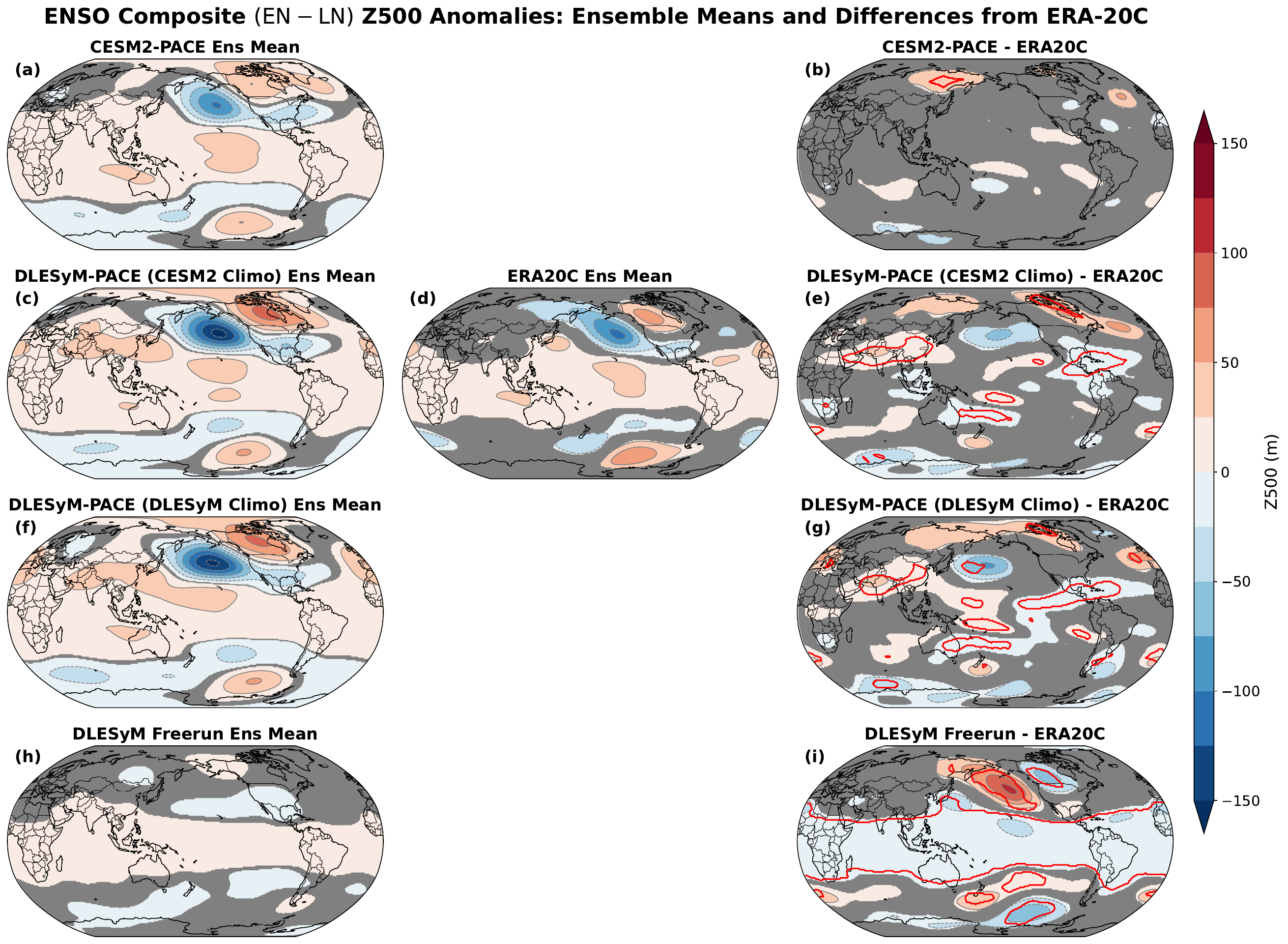}
\caption{ENSO composites (El Niño $-$ La Niña) of DJF Z500 (gpm) for (a) the ensemble mean of the 10 CESM2 Pacific Pacemaker (PACE) simulations, (c) DLESyM-PACE with CESM2 SST climatology, (d) ERA-20C, (f) DLESyM-PACE with DLESyM SST climatology, and (h) the DLESyM free-running experiment. 
Panels (b), (e), (g), and (i) show their respective differences relative to ERA-20C. 
For panels (a), (c), (d), (f), and (h), values not significant at the 5\% confidence level based on a two-sided $t$-test are shaded in gray. 
For panels (b), (e), (g), and (i), areas without shading indicate regions where the observed composite lies below the 5th percentile or above the 95th percentile of the bootstrapped ENSO composites. 
Red contours denote regions where the ERA-20C composite lies outside the range of all individual simulations.
}
\label{fig:mean response}
\end{figure}

The evaluation framework follows a methodology comparable to that used in Deser et al. (2017)\cite{Deser2017} (their figure 3a-c). The ensemble mean of the 10 pacemaker composites from CESM2 and DLESyM provides a robust estimate of each model’s response to ENSO, based on 200 El Niño events from each model, and 180 and 170 La Niña events from CESM2 and DLESyM, respectively, over the 1900-2009 period (Table \ref{enso_events}). This approach reveals a statistically significant Z500 response across most extratropical regions in both pacemaker experiments (Fig.~\ref{fig:mean response}a,c,f).

When forced with identical observed tropical SST anomalies, DLESyM’s atmospheric component produces a spatially realistic but significantly amplified ENSO teleconnection compared to the CESM2 benchmark. The general spatial pattern of high and low Z500 anomalies is consistent between the two models. However, DLESyM-PACE exhibits a stronger response, with negative anomalies over the North Pacific approximately 50m deeper and positive height anomalies over the Arctic and Canada about 50m higher than in CESM2 (Fig.\ref{fig:mean response}a,c,f). In contrast, the ENSO signal in the DLESyM free-running simulation is substantially weaker due to its weak ENSO variability (Figure 1c), highlighting the critical role of the pacemaker forcing in eliciting this strong extratropical response (Fig.\ref{fig:mean response}h).

Relative to the observed ENSO composite, DLESyM-PACE ensembles overestimate the amplitude of the extratropical response with stronger positive anomalies over Canada and more intense negative anomalies over the North Pacific than observed (Fig.\ref{fig:mean response}d).

While CESM2-PACE exhibits smaller biases (Fig.\ref{fig:mean response}b), DLESyM-PACE shows similar spatial biases but with stronger anomalies, particularly a high-pressure center shifted eastward toward the North Atlantic and exaggerated low anomalies over the North Pacific (Fig.\ref{fig:mean response}e,g). The deepened North Pacific low represents the largest magnitude bias in the DLESyM-PACE simulations. The weaker response in the free-running DLESyM across the Pacific–North American (PNA) sector further underscores that its intrinsic ENSO is too weak to drive a realistic teleconnection (Fig.\ref{fig:mean response}i).

\begin{figure}[!ht]
\centering
\includegraphics[width=\linewidth]{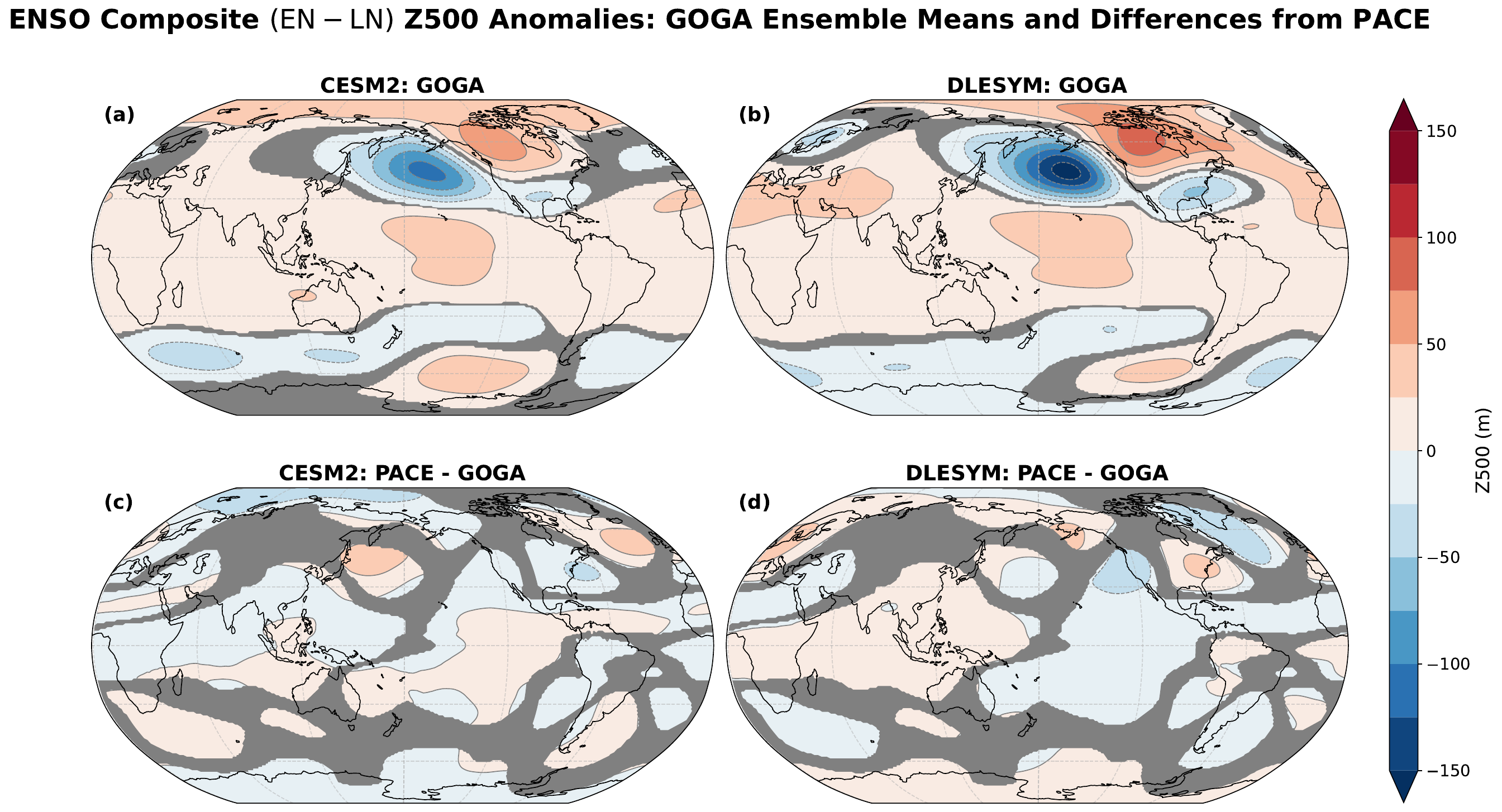}
\caption{ENSO composites (El Niño $-$ La Niña) of DJF Z500 (gpm) for prescribed global SST (GOGA) experiments using (a) CESM2 and (b) DLESyM. Difference maps showing the contribution from freely evolving SSTs outside the pacemaker region, calculated as PACE minus GOGA, for (c) CESM2 and (d) DLESyM (CESM2 Climo). In panels (a) and (b), gray shading indicates regions where the composite is not significant at the 5\% confidence level based on a two-sided t-test. In panels (c) and (d), gray shading indicates where the difference between the PACE and GOGA composites is not statistically significant at the 5\% confidence level.
}
\label{fig:pace goga diff response}
\end{figure}

To investigate the source of this amplified response, we performed an additional experiment to isolate the contribution of freely evolving SSTs outside the pacemaker region. By subtracting the response from an AMIP-style experiment with globally prescribed ERSSTv5 SSTs (GOGA) from the pacemaker (PACE) response, we can identify the impact of the coupled ocean dynamics outside the tropical Pacific. This diagnostic reveals that the primary teleconnection pattern is overwhelmingly driven by the tropical Pacific forcing in both models as shown in the similarity of pattern between Figure \ref{fig:pace goga diff response}a,b and Figure \ref{fig:mean response}a,c.

However, the GOGA experiment also suggests that DLESyM's amplification bias is an intrinsic property of its atmospheric model rather than an artifact of coupled ocean dynamics in the pacemaker runs. Although the GOGA simulations show a more consistent spatial pattern between CESM2 and DLESyM, the overestimation in the amplitude of DLESyM’s extratropical response remains. The SSTs outside of the pacemaker region do modulate the response regionally. In CESM2 (Fig. \ref{fig:pace goga diff response}c), meanwhile, they elongate the low anomaly further inland over East Russia, which aligns more closely with ERA20C (Fig. \ref{fig:mean response}e). In DLESyM (Fig. \ref{fig:pace goga diff response}d), they reduce the magnitude of the North Pacific low but strengthen the low over the eastern US, while also dampening the Canadian high, which helps reduce the overall bias shown in ERA20C (Fig. \ref{fig:mean response}e). Nevertheless, the persistent amplification points to a fundamental difference in the atmospheric model's sensitivity to tropical forcing

\subsection*{Evaluating DLESyM's internal variability}
Another crucial step in evaluating the model's forced response to ENSO is to determine if its internal atmospheric variability is realistic. The goal is to isolate true biases in the model's learned physical relationship from discrepancies that could arise simply from an unrealistic simulation of random atmospheric ``noise.'' As established by the methodology of Deser et al. (2017), if a model's internal variability is found to be in good agreement with observations, then any significant deviation where the observed composite lies outside the model's range of possibilities can be confidently attributed to a bias in its forced response. To perform this critical assessment, we create maps of the 90\% Confidence Intervals (CIs) for both the simulated and observed ENSO composites. These CIs are derived from the distribution of 2,000 bootstrapped samples, calculated at each grid point as the range between the 5th and 95th percentiles. The resulting CI map provides a spatial signature of uncertainty, where its magnitude, or ``spread,'' serves as a measure of the impact of internal variability. 

\begin{figure}[!ht]
\centering
\includegraphics[width=\linewidth]{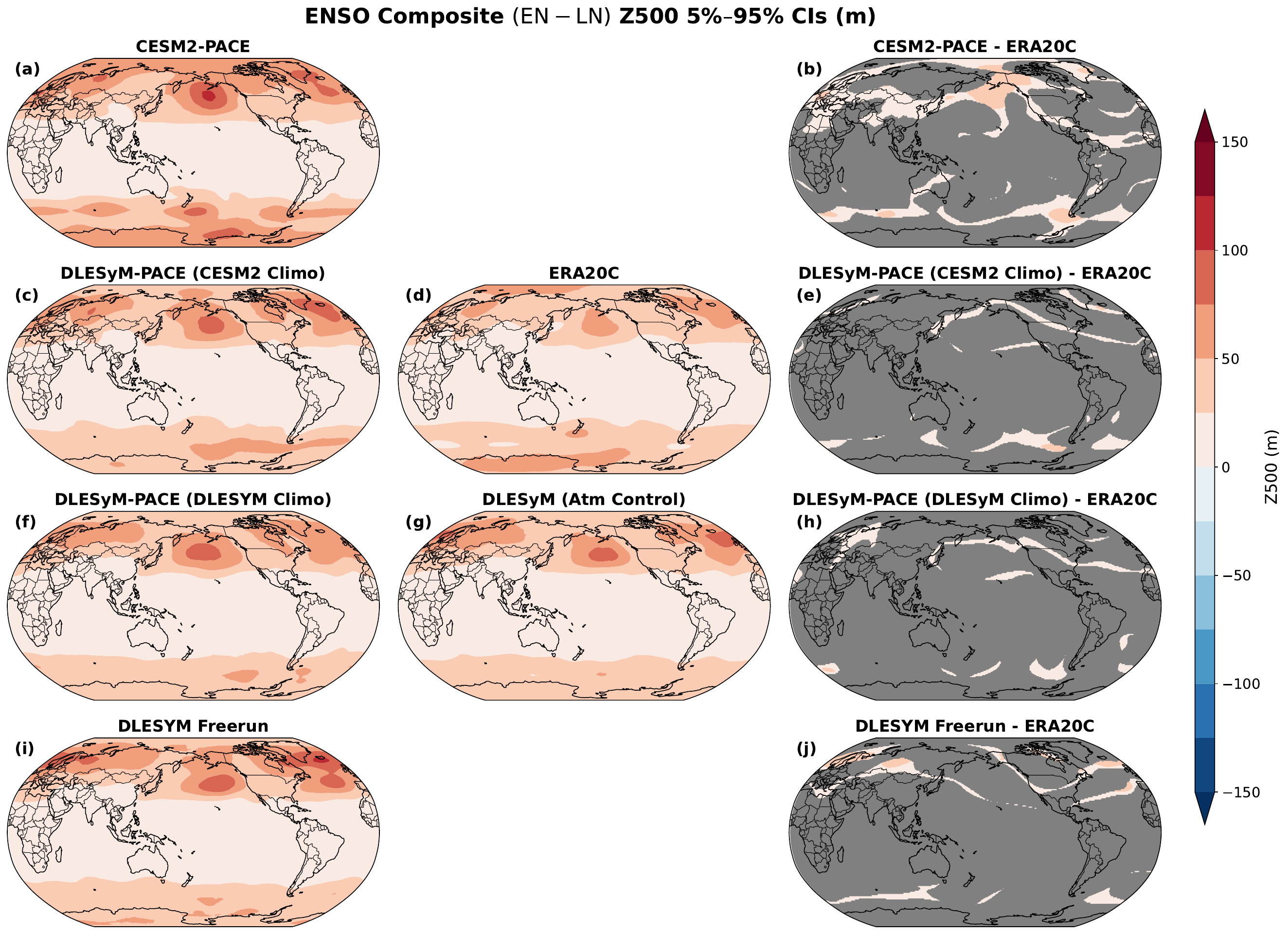}
\caption{The 5\%–95\% confidence intervals (CIs; gpm) of the Z500 ENSO composites for (a) the CESM2 Pacific Pacemaker (PACE) simulations, (c) DLESyM-PACE with CESM2 SST climatology, (d) ERA-20C, (f) DLESyM-PACE with DLESyM SST climatology, (g) the DLESyM atmospheric control experiment, and (i) the DLESyM free-running experiment. 
Panels (b), (e), (h), and (j) show their respective differences relative to ERA-20C. 
Gray shading indicates regions where the observed composite lies within the spread of values from the individual simulations. 
CIs are derived from 2000 bootstrapped samples for both the model simulations and ERA-20C.}
\label{fig:CI}
\end{figure}

Overall, the CIs for both the CESM2 (Fig. \ref{fig:CI}a) and DLESyM-PACE experiments (Fig. \ref{fig:CI}c,f) show a strong correspondence with the observed uncertainty (Fig. \ref{fig:CI}d) in both spatial pattern and magnitude.  Furthermore, the difference maps (Fig. \ref{fig:CI}b,e,h,j) allow for a more rigorous diagnosis of model performance by highlighting regions where the observed CI falls entirely outside the range of CIs from the individual ensemble members. In these maps, CESM2 (Fig. \ref{fig:CI}b) displays an overestimation of internal variability in the Arctic extending toward Alaska and in the southeast Pacific, with CI values exceeding the observed range by 25-50 m.

In contrast, the DLESyM-PACE simulations (Fig. \ref{fig:CI}e,h) exhibit a closer agreement with the observed variability, showing markedly fewer areas of significant discrepancy. This strong fidelity in simulating the magnitude and structure of internal atmospheric noise is critical, as it provides a robust foundation for the previous bias analysis. Because the "noise" characteristics are realistic in DLESyM, we can confidently attribute significant differences between its mean composite and the observed composite to biases in the model’s forced response. Furthermore, a comparison between the DLESyM-PACE CI maps and the CI derived from an atmosphere-only control run (Fig. \ref{fig:CI}g) confirms that this uncertainty is almost entirely rooted in internal atmospheric dynamics, a result consistent with the findings of Deser et al. 2017.

\subsection*{DLESyM atmospheric response to Different Pacemaker SST Climatology}
\begin{figure}[!ht]
\centering
\includegraphics[width=\linewidth]{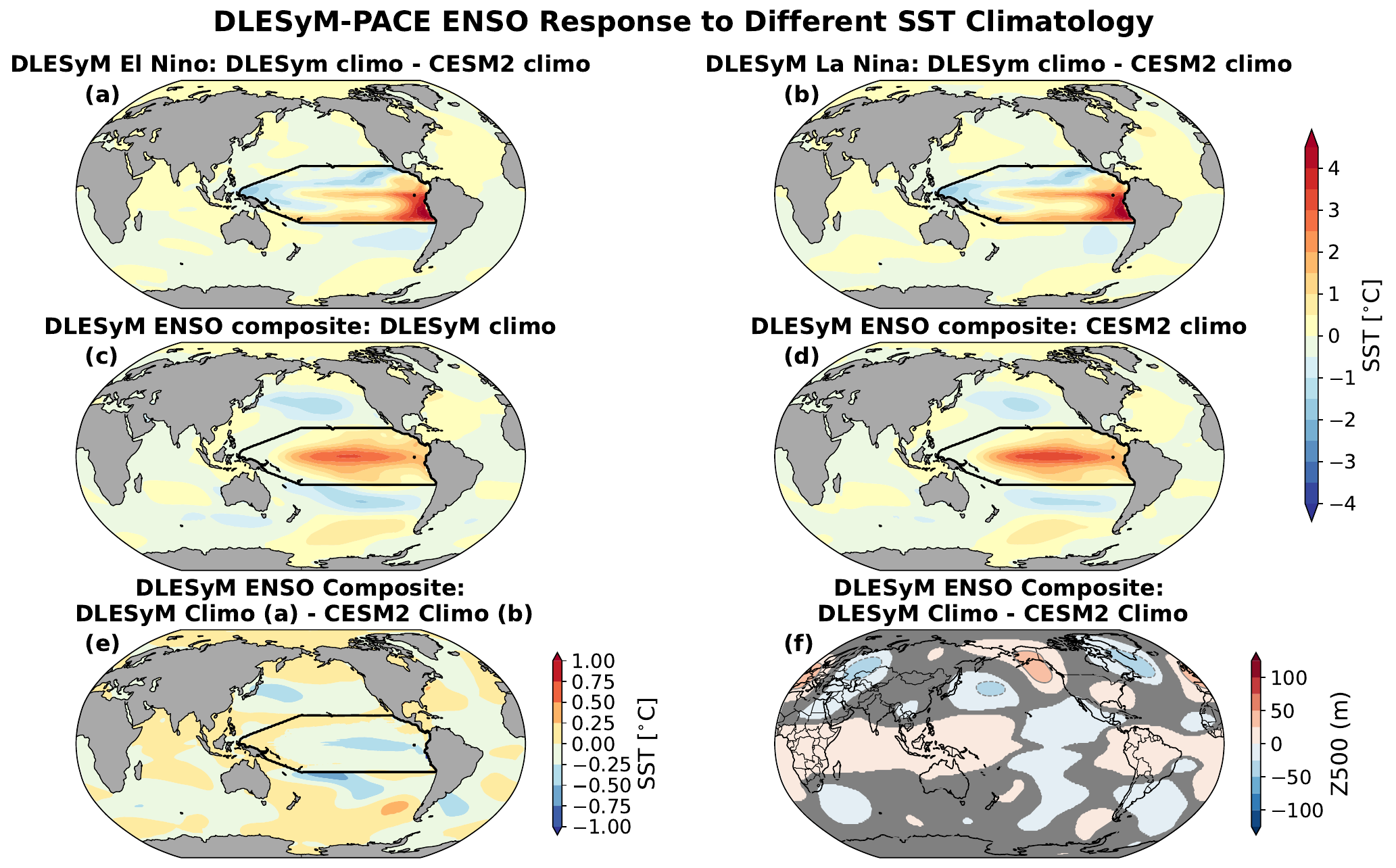}
\caption{DLESyM Pacific Pacemaker experiments comparing two different SST climatologies: one using the CESM2 climatology and the other using the DLESyM climatology within the Pacific region outlined by the black contour. 
(a) Ensemble-mean SST difference during El~Niño (EN) between the DLESyM and CESM2 climatologies; 
(b) same as (a) but for La~Niña (LN); 
(c) ENSO composite (EN--LN) of SST using the DLESyM climatology; 
(d) ENSO composite (EN--LN) of SST using the CESM2 climatology; 
(e) difference of ENSO composites between (c) and (d); 
(f) difference in the DLESyM Z500 ENSO response between the DLESyM and CESM2 climatologies (corresponding to Fig.~\ref{fig:mean response}c~--~\ref{fig:mean response}f). 
Values not significant at the 5\% confidence level based on a two-sided $t$-test are shaded in gray.
}
\label{fig:sst diff}
\end{figure}

The different tropical Pacific climatological SST in the two pacemaker experiments (DLESym Climo vs. CESM2 Climo ) yields slightly different El Niño and La Niña composite anomalies (Figure \ref{fig:sst diff}a,b). In the DLESyM-climatology run (Fig.\ \ref{fig:sst diff}c), the ENSO SST anomalies are weaker across the tropical Pacific, compared to the CESM2-climatology run (Fig.\ \ref{fig:sst diff}d). The DLESyM Climo case (Fig.\ \ref{fig:sst diff}a,b) shows a larger area of stronger negative SST anomalies along the South Pacific Convergence Zone (SPCZ) and smaller area of positive SST anomalies in the Kuroshio current. In other words, compared to the CESM2 case, the DLESyM-climatology experiment has an intensified cooling over the western boundary current and subtropical Pacific during El Niño (Fig.\ \ref{fig:sst diff}a).

These regional SST differences have important implications for atmospheric Rossby wave sources. In the Kuroshio region, the enhanced negative SST anomalies in DLESyM imply a stronger local heat sink and reduced convection, which alters the wind stress curl in the subtropical North Pacific. This is consistent with Qiao et al. (2023)\cite{Qiao2023ENSORelatedIV}, who showed that ENSO-driven wind curl anomalies in the western North Pacific excite westward-propagating baroclinic Rossby waves along the subtropical countercurrent, thereby modulating the Kuroshio recirculation strength. By analogy, the pronounced Kuroshio cooling in the DLESyM run likely intensifies this Rossby wave response. The result is a deeper 500-hPa trough over the western Pacific under the DLESyM climatology (Fig.\ \ref{fig:sst diff}f), compared to the CESM2 case.

Similarly, the SPCZ anomalies in DLESyM suggest a modified tropical–extratropical teleconnection. van der Wiel et al. (2015)\cite{vanderWiel2015ADF} describe how ENSO perturbs the SPCZ via changes in the upper-level westerly duct. They show that La Niña–like forcing (enhanced convection over the eastern Indian Ocean) expands the westerly duct and shifts the SPCZ westward, whereas El Niño heating in the central Pacific contracts the duct and shifts the SPCZ eastward. The stronger negative SST anomalies along the SPCZ in the DLESyM run imply a suppressed subtropical heat source, analogous to an amplified El Niño regime. In this scenario the westerly duct would be contracted and the Rossby wave source shifted eastward. This altered wave source plausibly steers a different upper-tropospheric response. Indeed, the Z500 composites (Fig.\ \ref{fig:sst diff}d) reveal that the DLESyM-climatology experiment produces a more pronounced trough–ridge dipole downstream of the SPCZ cooling. 

\subsection*{ENSO-induced Blocking Characteristics}
The stronger and spatially biased Z500 responses described above may also influence the occurrence of rare weather patterns such as atmospheric blocking. To examine this connection, we evaluate the seasonal blocking frequency, defined as the percentage of blocked days per season (DJF) at each grid point. The analysis focuses on two regions strongly affected by ENSO teleconnections: the North Pacific sector (120°E–120°W) and the Euro-Atlantic sector (60°W–60°E). We compare the blocking frequency, size, and intensity during El Niño and La Niña events between the ERA20C reanalysis and the pacemaker experiments.

\begin{figure}[!ht]
\centering
\includegraphics[width=16cm]{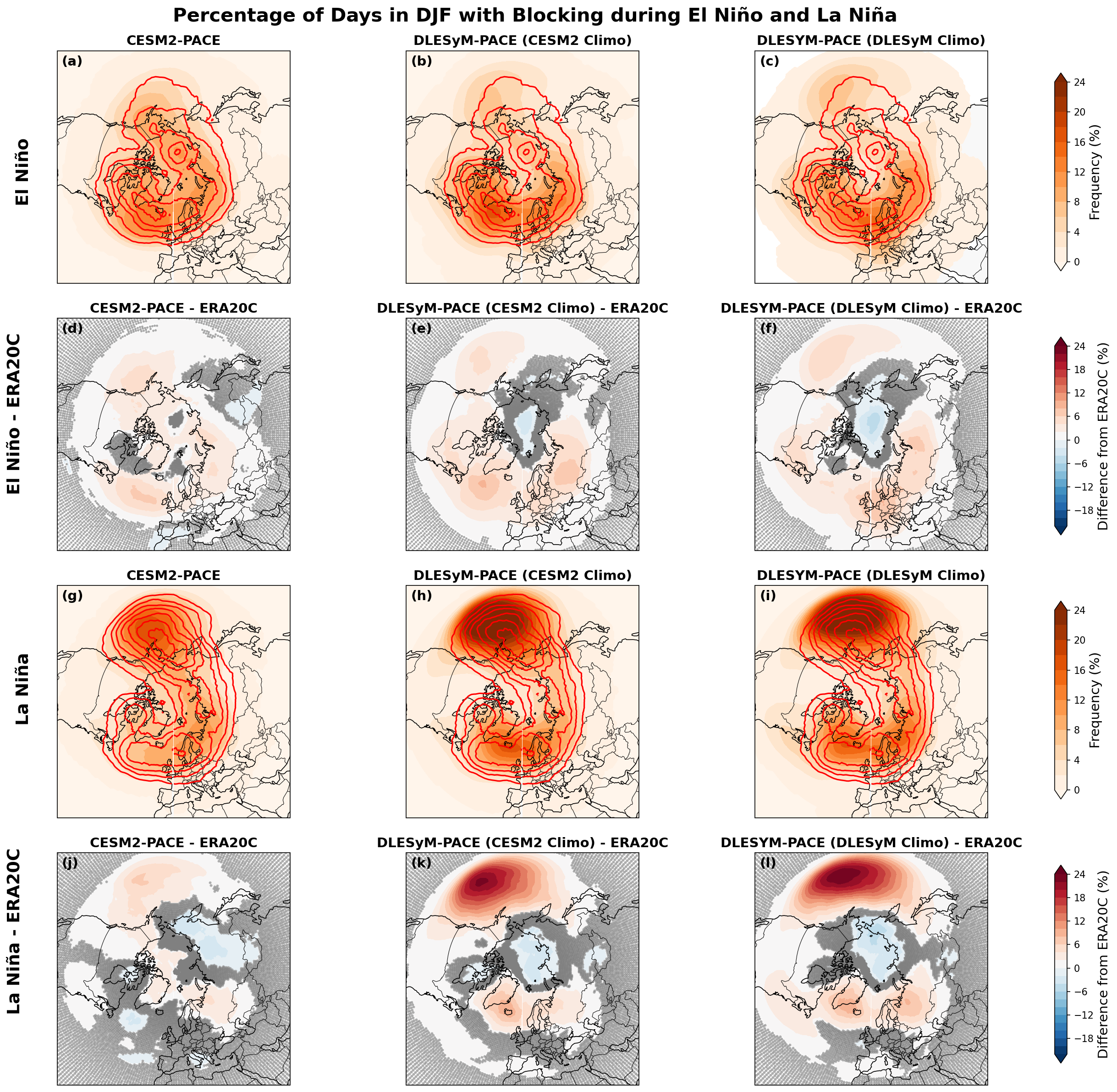}   
\caption{Boreal Winter (DJF) Blocking frequency of the Pacific Pacemaker experiments for CESM2, DLESyM with CESM2 SST climatology, and DLESyM with its own SST climatology during El~Niño \big((a)–(c)\big) and La~Niña \big((g)–(i)\big). Solid red contour indicate the climatological blocking of ERA20C with the outermost contour equal to 2\%, increasing inwards by a 2\% step. Panels (d)–(f) and (j)–(l) show the corresponding differences relative to ERA-20C for El~Niño and La~Niña, respectively. 
Stippling indicates regions not significant at the 5\% confidence level based on a two-sided $t$-test.
}
\label{fig:blocking anomaly}
\end{figure}

During El~Niño events (Fig.~\ref{fig:blocking anomaly}a–c), the CESM2-PACE simulation reproduces the observed blocking distribution reasonably well, showing enhanced blocking over Alaska and the Bering Strait consistent with ERA20C (red contours). This means that the spatial pattern of blocking during El Niño events can be attributed to the tropical forcing. The magnitude of blocking differs by roughly 4\% between CESM2-PACE and ERA20C (Fig.~\ref{fig:blocking anomaly}d), suggesting a contribution from non-tropical Pacific forcing. In contrast, the DLESyM-PACE response exhibits larger discrepancies with reanalysis in both pattern and magnitude. The DLESyM simulations show a southward-shifted blocking center over the North Pacific and more frequent blocking over northeastern Canada and Scandinavia (Fig.~\ref{fig:blocking anomaly}e–f), implying the influence of coupled ocean-atmospheric processes outside the Pacific pacemaker region, as well as potential atmospheric model biases.

During La~Niña events (Fig.~\ref{fig:blocking anomaly}g–i), all pacemaker experiments simulate increased blocking across the North Pacific compared to ERA20C, with DLESyM-PACE showing anomalies exceeding 20\% and the blocking center extending farther south. Moderate increases are also observed over Greenland and western Russia, with DLESyM exhibiting anomalies over 8\%, whereas CESM2-PACE shows a decrease of about 4\% over Greenland and an increase of roughly 4\% over western Russia. 
% Nevertheless, DLESyM continues to display a persistent high-latitude bias with reduced blocking frequency in polar regions, suggesting that background circulation biases may influence how ENSO teleconnections modulate blocking.

\begin{figure}[!ht]
\centering
\includegraphics[width=14cm]{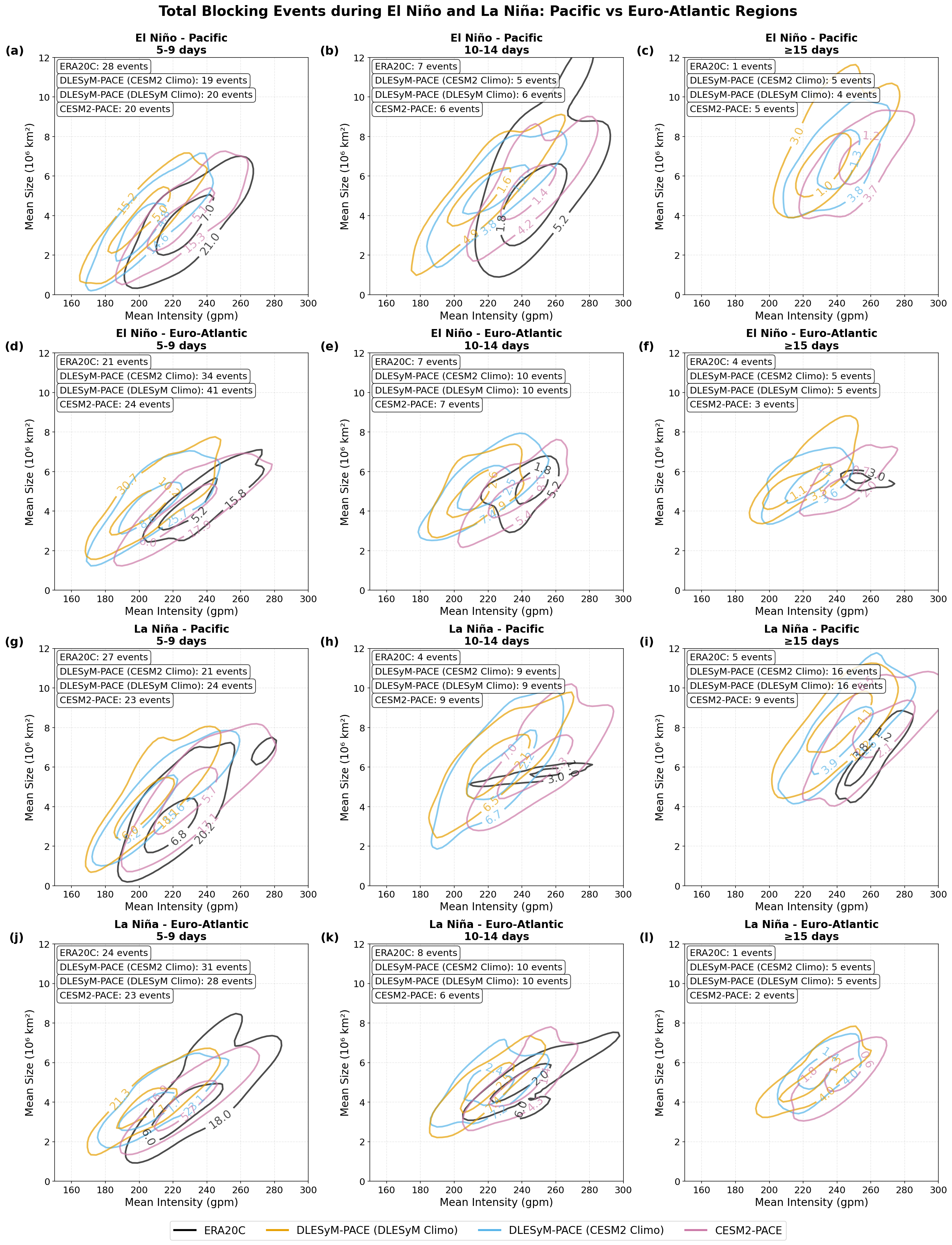}
\caption{Kernel density estimation (KDE) plots of blocking events from different Pacific Pacemaker experiments and ERA-20C (see legend at the bottom) for three categories of blocking duration: 5–9~days \big((a), (d), (g), (j)\big), 10–14~days \big((b), (e), (h), (k)\big), and $\geq$15~days \big((c), (f), (i), (l)\big). 
Panels (a)–(f) correspond to El~Niño events, and panels (g)–(l) correspond to La~Niña events during boreal winter. 
Each KDE is evaluated as a joint distribution of blocking size and intensity over two regions: the Pacific sector (120°E–120°W, 30°–90°N; panels~a–c and~g–i) and the Euro-Atlantic sector (60°W–60°E, 30°–90°N; panels~d–f and~j–l). 
Contours denote the 25th and 75th percentiles of total blocking events, with contour labels indicating the number of events within each percentile range for the corresponding duration category across all El~Niño or La~Niña events from 1900–2009. 
Text in each panel legend indicates the total number of blocking events detected for that experiment.
 }
\label{fig:blocking stats}
\end{figure}

The large bias in blocking frequency may result from either more frequent short-duration blocking events or modest increases in long-duration events exceeding 15 days. Since deep learning models often struggle to reproduce extreme events, it is also useful to examine the detailed blocking characteristics. Figure~\ref{fig:blocking stats} shows the relationship between blocking event duration, size, and intensity, with the 25th and 75th percentile contours representing the distribution of blocking events in both the North Pacific and Euro-Atlantic sectors during El Niño and La Niña phases.

In the North Pacific during El Niño (Fig.~\ref{fig:blocking stats}a–c), both CESM2 and DLESyM simulate fewer short-lived (5–9 day) blocking events compared to ERA20C (approximately 20 vs. 28 events), but they generate a higher number of longer-duration blocking episodes. This leads to an overall high bias in blocking frequency, primarily due to these extended events. In contrast, within the Euro-Atlantic sector (Fig.~\ref{fig:blocking stats}d–f), the increased blocking frequency relative to ERA20C arises mainly from shorter 5–9 day events, with CESM2-PACE showing 24 events and DLESyM over 30, compared to 21 in ERA20C.

During La Niña, the strong positive blocking anomalies over the North Pacific in the DLESyM models (Fig.~\ref{fig:blocking anomaly}k,l) can be attributed to roughly five additional events lasting 10–14 days and nine more events exceeding 15 days, relative to ERA20C. The strong negative Z500 ENSO composite anomaly in DLESyM-PACE likely influences the tail of the blocking-duration distribution, enhancing extreme event frequency. In the Euro-Atlantic region, the positive bias in DLESyM is weaker but manifests as a moderate increase across all three duration categories.

Regarding blocking size and intensity, both CESM2 and DLESyM show correlations between these metrics similar to those found in ERA20C, and the overall blocking size distribution is realistically represented. The climatological ENSO response in blocking also aligns closely with ERA20C, although biases persist primarily in blocking intensity. DLESyM tends to underestimate high-intensity blocking events, typically producing amplitudes 30–40~gpm lower than ERA20C. This bias is somewhat reduced when DLESyM is forced with the CESM2 SST climatology, whereas CESM2 itself exhibits closer agreement with ERA20C in both spatial extent and intensity.

\section*{Discussion}

This study presents the application of the Pacific pacemaker framework to a deep learning ocean coupled climate emulator, DLESyM \cite{CresswellClay2024ADL} to evaluate its forced atmospheric response to ENSO. Our pacemaker experiments reveal that DLESyM’s atmospheric component successfully captures the canonical spatial pattern of the wintertime extratropical teleconnection, but with a significantly amplified magnitude compared to both observations and the CESM2 benchmark. This type of amplitude bias is a well-documented issue in traditional general circulation models \cite{Chen2020ModelBI}, but its characterization in a deep learning emulator provides new insight into the capabilities and limitations of these models. We find that DLESyM realistically simulates the magnitude and structure of internal atmospheric variability. Following the diagnostic framework of Deser et al. 2017 \cite{Deser2017}, this high fidelity in the atmospheric variability provides strong evidence that the amplified teleconnection is a genuine bias in the model’s forced response, not an artifact of sampling variability.

By forcing the same deep learning atmosphere with different mean-state sea surface temperatures, we demonstrate that regional biases, particularly in the Kuroshio Current and South Pacific Convergence Zone, directly alter the resulting Rossby wave propagation and extratropical circulation response. This finding provides a direct, mechanistic link between mean-state errors and teleconnection biases, in contrast to studies that have suggested a minimal impact of model biases on ENSO teleconnections \cite{tyrrell2021minimal}.

The consequences of this amplified mean response extend to the simulation of extreme weather. DLESyM overestimates the frequency and duration of atmospheric blocking events, particularly during La Niña. However, it simultaneously underestimates their peak intensity which is one of the challenges AI models face with predicting the extreme \cite{Hua2025PerformanceOT}. It suggests the model correctly learns the conditions conducive to blocking but fails to reproduce the highest-amplitude, rare events that lie at the edge of its training distribution, a problem often referred to as forecasting "gray swan" events \cite{sun2025can}.

Overall, this study demonstrates both the promise and the current limitations of deep learning coupled climate emulators in simulating key tropical-extratropical interactions. Indeed, even recent emulators that couple 3D ocean and atmosphere models, such as SamudrACE \cite{Duncan2025SamudrACE}, still exhibit weak ENSO variability, reinforcing the need for such diagnostic experiments. While the emulator DLESyM used here can capture key dynamical responses with high fidelity, it also exhibits higher sensitivity to tropical forcing than its dynamical model counterpart, and difficulty with simulating key characteristics of rare extremes, such as the duration of atmospheric blocking events. This underscores the need for targeted improvements, which can benefit from the pathway that this work establishes, that of using the pacemaker framework for diagnosing and ultimately reducing systematic teleconnection and extreme-event biases. As AI-based climate models continue to develop, such physically grounded assessments will be essential for building confidence in their use for future climate projections and risk analysis.
\section*{Methods}

This section details the observational datasets used for model forcing and evaluation, the climate models and experimental configurations employed, and the statistical techniques used for analysis. The experimental design is structured to isolate the atmospheric response to observed El Niño–Southern Oscillation (ENSO) forcing within a novel deep learning ocean atmosphere coupled emulator and to disentangle the roles of tropical forcing, internal atmospheric variability, and background climatology in shaping extratropical teleconnections.

\subsection*{Reanalysis Data}

\subsubsection*{Sea Surface Temperature (SST)}
For prescribing tropical Pacific SST anomalies in the pacemaker experiments and for identifying historical ENSO events, we utilize the monthly mean NOAA Extended Reconstructed Sea Surface Temperature, version 5\cite{huang2017extended} (ERSSTv5) dataset, which has a $2^{\circ} \times 2^{\circ}$ resolution. The analysis period covers the 110 years from 1900 to 2009.

\subsubsection*{Atmospheric Circulation}
The primary variable for evaluating extratropical teleconnections is the 500 hPa geopotential height (Z500). To provide a consistent, long-term observational benchmark, we use the mean of the ensemble reanalysis of the twentieth century (ERA-20C\cite{poli2016era}) from the European Centre for Medium-Range Weather Forecasts (ECMWF). All reanalysis data are regridded to a common resolution for comparison with model output.

\subsection*{Model Descriptions and Experimental Design}

This study performs a hierarchical set of experiments to examine ENSO teleconnections in an AI-based vs. a dynamical climate model: The dynamical model is the Community Earth System Model version 2 (CESM2\cite{danabasoglu2020community}) and the AI-based climate emulator is the novel Deep Learning Earth System Model (DLESyM\cite{CresswellClay2024ADL}).

\subsubsection*{The Community Earth System Model version 2 (CESM2)}
We use simulations from the publicly available CESM2 project as a benchmark for state-of-the-art climate model performance. CESM2 is a fully coupled global climate model with interactive atmosphere, ocean, land, and sea ice components, featuring a nominal 1-degree horizontal resolution in both the atmosphere and ocean. 

\paragraph{The CESM2 Large Ensemble (CESM2-LENS2)} 
This 50-member ensemble of fully coupled, freerunning simulations serves as our primary reference for internally generated ENSO variability and its associated teleconnections in a conventional model framework. For this study, we utilize members 1-10. Each member is forced with historical and projected radiative forcings but evolves from a slightly perturbed initial condition, initialized from a specific segment of a 1400-year pre-industrial control simulation.

\paragraph{The CESM2 Global Prescribed SST AMIP (CESM2-GOGA)} This experiment consists of an ensemble of atmosphere-only simulations where the atmospheric component of CESM2 is forced with historically observed global SSTs (ERSSTv5) and sea ice concentrations (HadISST1 and OISSTv2). Unlike the coupled PACE experiment, the GOGA setup eliminates the influence of coupled ocean-atmosphere dynamics entirely. Its primary role in our analysis is to serve as a diagnostic tool. By comparing the ENSO response in CESM2-PACE to that in CESM2-GOGA, we can isolate and quantify the contribution of the freely evolving, coupled ocean outside the tropical Pacific pacemaker region.

\paragraph{The CESM2 Pacific Pacemaker (CESM2-PACE)} This experiment follows the "pacemaker" protocol designed to isolate the causal impact of observed tropical SST variability. In this 10-member ensemble, SST anomalies within the tropical Pacific (10°S–10°N, linearly tapering to 20°S–20°N) are nudged toward observations from ERSSTv5. This methodology forces the model's ENSO evolution to follow the observed historical timeline while allowing the rest of the global climate system to remain fully coupled and evolve freely. This ensemble provides the benchmark for the "true" forced atmospheric response to observed ENSO in a GCM. 

\subsubsection*{The Deep Learning Earth System Model (DLESyM)}
DLESyM \cite{CresswellClay2024ADL} is a novel, hybrid coupled climate emulator that pairs a deep learning (DL) atmospheric model with a traditional, simplified dynamical ocean model. The atmospheric component is an AI-based emulator trained on reanalysis data. The two components are coupled asynchronously: the atmospheric model is integrated forward for a 96-hour (4-day) period using a fixed SST field. The time-averaged surface fluxes (e.g., wind stress, pressure) from this period are then used to force the ocean model, which generates an updated SST field for the subsequent 96-hour atmospheric integration. All experiments involving DLESyM were conducted using model checkpoints and utilities from NVIDIA's Earth2Studio framework \url{https://github.com/NVIDIA/earth2studio}.

\paragraph{DLESyM Freerunning Ensemble (DLESyM-Freerun)} To establish the baseline behavior of the emulator, we generated a 10-member freerunning ensemble. Each member was initialized from a different, randomly selected atmospheric state from the ERA5 reanalysis. The model was then run for a spin-up period of 80 years to allow the coupled system to reach a stable, internally consistent climatology. Following this spin-up, we recorded a 110-year simulation from each member (mimicking the period 1900-2009) to characterize the model's mean state, its representation of ENSO, and its baseline extratropical teleconnections.

\paragraph{DLESyM Atmospheric Control Ensemble (DLESyM-Atmos)} To quantify the internal variability generated solely by the DL atmospheric component, we conducted an atmosphere-only control experiment. This 10-member ensemble was forced with a prescribed, repeating seasonal cycle of SST derived from the long-term climatology of the DLESyM-Freerun ensemble. By design, this experiment contains no interannual ocean variability (i.e., no ENSO). The spread across this ensemble provides a clean measure of the model's internal atmospheric "noise," which is a crucial diagnostic for interpreting the pacemaker results. This approach is directly analogous to the atmospheric control simulations used in foundational studies to isolate the contribution of unpredictable atmospheric variability from the forced response.

\paragraph{The DLESyM Prescribed SST AMIP (DLESyM-GOGA)} In this configuration, the atmospheric component of DLESyM is run without coupling to its internal ocean model. Instead, it is forced with the same prescribed historical global SSTs from the ERSSTv5 dataset, using the same approach as in Meng et al. (2025) \cite{meng2025deep}. This setup allows us to evaluate the atmospheric model's response to observed global SST variability in the absence of any coupled ocean feedbacks. Similar to its CESM2 counterpart, the DLESyM-GOGA ensemble is crucial for attributing the sources of teleconnection biases by isolating the impact of the simplified coupled ocean model operating outside the pacemaker region.

\paragraph{DLESyM Pacific Pacemaker Ensemble (DLESyM-PACE)} This is the main experiment of our study, representing one of the first application of the pacemaker methodology to a deep learning climate emulator. The nudging mask in the tropical Pacific was designed to be geographically identical to that used in the CESM2-PACE experiments. Monthly mean SST anomalies from ERSSTv5 were linearly interpolated to a 48-hour interval and applied as a nudging tendency at each step of the ocean model's integration within the 96-hour coupling cycle. For each ensemble member, the model was initialized from a different atmospheric state from January 1st of various years in the 1950s. Nudging of tropical Pacific SST anomalies to observed ERSSTv5 values began immediately, using observations from 1890. A 10-year adjustment period (1890–1899) was allowed for the atmosphere and ocean to equilibrate to the pacemaker forcing before the primary analysis period of 1900–2009 was recorded.

The DLESyM-PACE experiments are conducted under two distinct background climatologies, allowing us to move beyond merely identifying teleconnection errors and toward causally attributing their sources. Standard pacemaker comparisons (e.g., DLESyM-PACE vs. CESM2-PACE) can demonstrate that models respond differently to the same tropical forcing, but they cannot distinguish whether the differences arise from the models' representation of atmospheric physics or from biases in their background climate states.
\begin{enumerate}
\item \textbf{DLESyM-PACE (DLESyM Climo):} A 10-member ensemble where the nudged ERSSTv5 SST anomalies are added to the DLESyM's own native climatological mean SST, as derived from the DLESyM-Freerun ensemble.
\item \textbf{DLESyM-PACE (CESM2 Climo):} A second 10-member ensemble where the same nudged ERSSTv5 SST anomalies are added to the climatological mean SST from the CESM2-LENS2.
\end{enumerate}
By forcing the exact same DL atmospheric model to operate within two different mean states, this experimental framework directly isolates and quantifies the impact of the background climatology on the teleconnection response. Any difference in the extratropical response between the two PACE with different SST climatology can be causally attributed to the change in the background state, thus providing a clean answer to our second research question.

\subsection*{Analysis Techniques}

The analysis methodology is designed to be consistent with established best practices for studying ENSO teleconnections and their uncertainty, primarily following the framework of Deser et al. (2017) \cite{Deser2017}.

\subsubsection*{ENSO Event Identification and Compositing}
El Niño (EN) and La Niña (LN) events are identified for the period 1900–2009 using the Niño-3.4 index, calculated from the each experiment dataset as the area-averaged SST anomaly over $5^{\circ}$S–$5^{\circ}$N, $170^{\circ}$W–$120^{\circ}$W. Following the standard procedure outlined in Deser et al. (2017)\cite{Deser2017}, the monthly index is smoothed with a three-point binomial filter, and the November-January (NDJ) seasonal average is computed. An event is defined when this detrended NDJ index exceeds one standard deviation (for El Niño) or falls below negative one standard deviation (for La Niña). This procedure yields 21 El Niño and 18 La Niña events for the ERSSTv5 reanalysis. For the model climatologies plus the ERSSTv5 SST anomalies, there are 20 El Niño and 18 La Niña events for CESM2, and 20 El Niño and 17 La Niña events for DLESYM over the 1900–2009 period (Table \ref{enso_events}).

\begin{longtable}{ccccccc}
\label{enso_events}\\
\caption{El Niño and La Niña Events from ERSSTv5, CESM2, and DLESYM Climatologies}\\
\toprule
\textbf{Event Year} & \multicolumn{2}{c}{\textbf{ERSSTv5}} & \multicolumn{2}{c}{\textbf{CESM2}} & \multicolumn{2}{c}{\textbf{DLESYM}} \\
\cmidrule(lr){2-3} \cmidrule(lr){4-5} \cmidrule(lr){6-7}
& \textbf{El Niño} & \textbf{La Niña} & \textbf{El Niño} & \textbf{La Niña} & \textbf{El Niño} & \textbf{La Niña} \\
\midrule
\endfirsthead
\multicolumn{7}{c}%
{{\bfseries \tablename\ \thetable{} -- continued from previous page}} \\
\toprule
\textbf{Event Year} & \multicolumn{2}{c}{\textbf{ERSSTv5}} & \multicolumn{2}{c}{\textbf{CESM2}} & \multicolumn{2}{c}{\textbf{DLESYM}} \\
\cmidrule(lr){2-3} \cmidrule(lr){4-5} \cmidrule(lr){6-7}
& \textbf{El Niño} & \textbf{La Niña} & \textbf{El Niño} & \textbf{La Niña} & \textbf{El Niño} & \textbf{La Niña} \\
\midrule
\endhead
\bottomrule
\endfoot
1902-1903 & 1902-1903 & - & 1902-1903 & - & 1902-1903 & - \\
1903-1904 & - & 1903-1904 & - & 1903-1904 & - & 1903-1904 \\
1904-1905 & - & - & - & - & 1904-1905 & - \\
1905-1906 & 1905-1906 & - & 1905-1906 & - & 1905-1906 & - \\
1908-1909 & - & 1908-1909 & - & 1908-1909 & - & 1908-1909 \\
1909-1910 & - & 1909-1910 & - & 1909-1910 & - & 1909-1910 \\
1911-1912 & 1911-1912 & - & 1911-1912 & - & 1911-1912 & - \\
1913-1914 & 1913-1914 & - & - & - & 1913-1914 & - \\
1914-1915 & 1914-1915 & - & 1914-1915 & - & 1914-1915 & - \\
1916-1917 & - & 1916-1917 & - & 1916-1917 & - & 1916-1917 \\
1917-1918 & - & 1917-1918 & - & 1917-1918 & - & - \\
1918-1919 & 1918-1919 & - & 1918-1919 & - & 1918-1919 & - \\
1924-1925 & - & 1924-1925 & - & 1924-1925 & - & 1924-1925 \\
1925-1926 & 1925-1926 & - & 1925-1926 & - & 1925-1926 & - \\
1930-1931 & 1930-1931 & - & 1930-1931 & - & 1930-1931 & - \\
1933-1934 & - & 1933-1934 & - & 1933-1934 & - & 1933-1934 \\
1938-1939 & - & 1938-1939 & - & 1938-1939 & - & 1938-1939 \\
1940-1941 & 1940-1941 & - & 1940-1941 & - & 1940-1941 & - \\
1941-1942 & 1941-1942 & - & 1941-1942 & - & - & - \\
1942-1943 & - & 1942-1943 & - & 1942-1943 & - & 1942-1943 \\
1949-1950 & - & 1949-1950 & - & 1949-1950 & - & 1949-1950 \\
1955-1956 & - & 1955-1956 & - & 1955-1956 & - & 1955-1956 \\
1957-1958 & 1957-1958 & - & 1957-1958 & - & 1957-1958 & - \\
1963-1964 & 1963-1964 & - & 1963-1964 & - & - & - \\
1965-1966 & 1965-1966 & - & 1965-1966 & - & 1965-1966 & - \\
1968-1969 & - & - & - & - & 1968-1969 & - \\
1970-1971 & - & 1970-1971 & - & 1970-1971 & - & 1970-1971 \\
1972-1973 & 1972-1973 & - & 1972-1973 & - & 1972-1973 & - \\
1973-1974 & - & 1973-1974 & - & 1973-1974 & - & 1973-1974 \\
1975-1976 & - & 1975-1976 & - & 1975-1976 & - & 1975-1976 \\
1982-1983 & 1982-1983 & - & 1982-1983 & - & 1982-1983 & - \\
1986-1987 & 1986-1987 & - & 1986-1987 & - & 1986-1987 & - \\
1987-1988 & 1987-1988 & - & 1987-1988 & - & - & - \\
1988-1989 & - & 1988-1989 & - & 1988-1989 & - & 1988-1989 \\
1991-1992 & 1991-1992 & - & 1991-1992 & - & 1991-1992 & - \\
1994-1995 & 1994-1995 & - & 1994-1995 & - & 1994-1995 & - \\
1997-1998 & 1997-1998 & - & 1997-1998 & - & 1997-1998 & - \\
1998-1999 & - & 1998-1999 & - & 1998-1999 & - & 1998-1999 \\
1999-2000 & - & 1999-2000 & - & 1999-2000 & - & 1999-2000 \\
2002-2003 & 2002-2003 & - & 2002-2003 & - & 2002-2003 & - \\
2007-2008 & - & 2007-2008 & - & 2007-2008 & - & 2007-2008 \\
\end{longtable}

The atmospheric response is analyzed using composites for the boreal winter season for each experiment, defined as December-February (DJF). The one-month lag relative to the NDJ index accounts for the time required for the atmospheric Rossby wave response to tropical heating to fully establish. The final ENSO composite is calculated as the mean of all EN winters minus the mean of all LN winters. 

\subsubsection*{Statistical Significance and Uncertainty Estimation}
A challenge in evaluating teleconnections is that the single observed historical record is contaminated by internal atmospheric variability ("noise") that is unrelated to ENSO. The observed composite, therefore, represents only one plausible realization of the forced response. To account for this, we adopt a random sampling (bootstrapping) methodology as in Karamperidou et al. 2013\cite{karamperidou2014intrinsic} and McKenna et al. 2023\cite{McKenna2023TheIO} to quantify the uncertainty inherent in any finite sample of ENSO events. 

We generate 2,000 ``bootstrapped'' synthetic ENSO composites. Each synthetic composite is created by randomly sampling, with replacement, 21 winters from the pool of identified EN years and 18 winters from the pool of identified LN years. The difference between these two randomly sampled means forms one synthetic composite. This procedure is applied to the observational record and to the model simulations. The distribution of these 2,000 synthetic composites provides a robust estimate of the uncertainty in the ENSO response that arises from the finite sample of events and the random superposition of internal variability.

A model's teleconnection pattern is considered statistically robust if its ensemble-mean response is significantly different from zero. It is considered consistent with observations if the observed composite falls within the 5th–95th percentile range of the model's bootstrapped distribution. This approach treats the observed record not as a perfect target, but as a single, noise-contaminated realization from a broader distribution of possibilities, providing a more rigorous basis for model evaluation.

\subsubsection*{Blocking Identification and Metrics}
Blocking characteristics associated with ENSO phases are analyzed for the pacemaker experiments. Blocking events were detected using the ConTrack algorithm \cite{Steinfeld2019TheRO} (\url{https://github.com/steidani/ConTrack}
), following the general approach described by Mckenna et al. 2023 \cite{McKenna2023TheIO}. The method identifies blocks as large-scale anomalies in the 500-hPa geopotential height (Z500) field relative to a climatological baseline. For each experiment, daily Z500 anomalies were computed with respect to the 1980–2009 climatology using a 31-day running mean window. To distinguish significant departures from the background state, seasonal thresholds were determined as the 90th percentile of Z500 anomalies from 50-80 degrees north during DJF.

A grid point is considered blocked when its Z500 anomaly exceeds the DJF threshold and the event persists for at least five consecutive days. The blocking frequency at each grid point is then expressed as the percentage of days per season identified as blocked. In addition, ConTrack quantifies two supplementary diagnostics for each block: extent and intensity. The extent represents the total area enclosed by the Z500 anomaly contour that delineates the spatial footprint of the block, while intensity is defined as the mean Z500 anomaly within that region.

\section*{Data Availability}
The ERSSTv5 is available at \url{https://psl.noaa.gov/data/gridded/data.noaa.ersst.v5.html}. The  ERA-20C is available at: \url{https://gdex.ucar.edu/datasets/d626000/}
The CESM2-LE simulations are accessible online at \url{https://www.cesm.ucar.edu/community-projects/lens2}. The CESM2 Pacific Pacemaker simulations are available at \url{https://www.cesm.ucar.edu/working-groups/climate/simulations/cesm2-pacific-pacemaker}. The CAM6 Prescribed SST AMIP Ensembles (CEMS2-GOGA) are available at \url{https://www.cesm.ucar.edu/working-groups/climate/simulations/cam6-prescribed-sst}. The DLESym model output used in this analysis will be deposited by the authors for public use on Zenodo at 10.5281/zenodo.17636221. 

\section*{Code Availability}
DLESyM model checkpoint available at NVIDIA's Earth2Studio framework \url{https://github.com/NVIDIA/earth2studio}. ConTrack for block tracking and identification is available at \url{https://github.com/steidani/ConTrack}.

\bibliography{sample}

\section*{Acknowledgements}
This work is supported by US National Science Foundation grant \#AGS-2202663. Computing resources provided by NCAR's Computational and Information Systems Laboratory, sponsored by the National Science Foundation \cite{cisl2019cheyenne}, were used to perform the reported analysis. This is SOEST contribution no. XXXX (completed in press). 

\section*{Author contributions statement}
ZH: Data curation, Formal analysis, Investigation, Methodology, Software, Validation, Visualization, Writing – original draft, Writing – review \& editing.  CK: Conceptualization, Funding acquisition, Investigation, Methodology, Project administration, Resources, Writing – review \& editing. ZM: Data curation, Software, Writing – review \& editing. All authors reviewed and approved the final manuscript.

% \section*{Additional information}
% \textbf{Funding} This work was supported by US National Science Foundation grant \#AGS-2202663.

% \textbf{Competing interests} The authors declare no competing interests.

\end{document}